\documentclass[fleqn]{article}
\usepackage{espcrc2}
\usepackage{psfig}
\usepackage{graphicx}

\newcommand{\BR}{{\cal B}}
\newcommand{\jpc}{J^{PC}}
\newcommand{\jpsi}{J/\psi}
\newcommand{\psp}{\psi(2S)}
\newcommand{\pspp}{\psi(3770)}
\newcommand{\y}{Y(4260)}
\newcommand{\oc}{\omega\chi_{c1}}

\newcommand{\chicz}{\chi_{c0}}
\newcommand{\chico}{\chi_{c1}}
\newcommand{\chict}{\chi_{c2}}
\newcommand{\psipp}{\pi^+\pi^- J/\psi}
\newcommand{\jpsipp}{\pi^+\pi^-J/\psi}
\newcommand{\ppjpsi}{\pi^+\pi^-J/\psi}
\newcommand{\EE}{e^+e^-}
\newcommand{\pip}{\pi^+}
\newcommand{\pim}{\pi^-}
\newcommand{\piz}{\pi^0}
\newcommand{\pp}{\pi^+\pi^-}
\newcommand{\ppp}{\pi^+\pi^-\pi^0}
\newcommand{\ccb}{c\overline{c}}

\newcommand{\ddb}{D\overline{D}}
\newcommand{\ra}{\rightarrow}

\newcommand{\beq}{\begin{equation}}
\newcommand{\eeq}{\end{equation}}
\newcommand{\beqn}{\begin{eqnarray}}
\newcommand{\eeqn}{\end{eqnarray}}
\newcommand{\beqns}{\begin{eqnarray*}}
\newcommand{\eeqns}{\end{eqnarray*}}
\newcommand{\btbl}{\begin{table}}
\newcommand{\etbl}{\end{table}}
\newcommand{\btbu}{\begin{tabular}}
\newcommand{\etbu}{\end{tabular}}
\newcommand{\bbcol}{BABAR collaboration }
\def\eref#1{(\ref{#1})}
\def\Journal#1#2#3#4{{#1} {\bf #2} (#4) #3}

\def\NCA{Nuovo Cimento A}
\def\NPB{Nucl. Phys. B}
\def\PLB{Phys. Lett. B}
\def\PRL{Phys. Rev. Lett.}
\def\PRD{Phys. Rev. D}
\title{\boldmath The $\y$ as an $\oc$ molecular state}
\author{C.Z.~Yuan \address[IHEP]{Institute of High Energy Physics,
P.O.Box 918, Beijing 100049, China}
\thanks{Supported by 100 Talents Program of CAS (U-25)
and National Natural Science Foundation of China (10491303).},
P.~Wang\addressmark[IHEP], and X.H.~Mo\addressmark[IHEP]}

\date{\today}
\begin{document}

\begin{abstract}
It is suggested that the newly observed $\y$ by \bbcol is a
molecular state composed of an $\omega$ and a $\chico$. Both the
production and decay properties are discussed. A consequence for
this molecular state, $\y$, is that it decays into $\ppp\chico$
with similar rate to $\ppjpsi$. It is also expected that $\y\to
\piz\piz \jpsi$ is produced at half rate as $\y\to\jpsipp$. These
decay modes should be searched for in the $B$-factories using
initial state radiative return data and $B$ decay data as well.
\end{abstract}

\maketitle

\section{Introduction}

Recently, in studying the initial state radiation events, $\EE \to
\gamma_{ISR} \ppjpsi$ ($\gamma_{ISR}$: initial state radiation
photon) with 233~fb$^{-1}$ data collected around
$\sqrt{s}=10.58$~GeV, \bbcol observed an accumulation of events
near 4.26~GeV/$c^2$ in the invariant-mass spectrum of
$\ppjpsi$~\cite{babay4260}. The fit to the mass distribution
yields $125 \pm 23$ events with a mass of $4259\pm
8^{+2}_{-6}$~MeV/$c^2$ and a width of $88\pm
23^{+6}_{-4}$~MeV/$c^2$.

Since the resonance is produced in initial state radiation from
$\EE$ collision, its quantum number $\jpc=1^{--}$. However, this
new resonance seems rather different from the known charmonium
states with $\jpc=1^{--}$ in the same mass scale, such as
$\psi(4040)$, $\psi(4160)$, and $\psi(4415)$. Being well above the
$\ddb$ threshold, instead of decaying predominantly into $\ddb$,
the $Y(4260)$ shows strong coupling to $\ppjpsi$ final state. So
this new resonance does not seem to be a usual charmonium state
but rather an exotic. The strange properties exhibited by the
$Y(4260)$ have triggered many theoretical
discussions~\cite{zhusl,maiani,llanes,kou,close,liux,qiaocf}.

One suggestion is that the $Y(4260)$ is the first orbital
excitation of a diquark-antidiquark state
($[cs][\bar{c}\bar{s}]$)~\cite{maiani}. By virtue of this scheme,
the mass of such a state is estimated to be 4.28~GeV/$c^2$, which
is in good agreement with the observation. A crucial prediction of
the scheme is that the $Y(4260)$ decays predominantly into $D_s
\overline{D}_s$.

Zhu scrutinized many possible interpretations for the $\y$ and
excluded the possibility of its being a conventional $\ccb$ state,
a $\ddb$ or $\omega\jpsi$ molecule, or a glueball using the
available experimental information~\cite{zhusl}. As to the
four-quark hypothesis, it is disfavored by its small total width
and the non-observation of the $\ddb$ decay mode. The author
regarded a hybrid charmonium as the most plausible interpretation,
which was consistent with all the experimental information by then.
Some other work to explain the $\y$ as a hybrid were put forth
afterwards~\cite{kou,close}. In the light of the lattice inspired
flux-tube model, the calculation shows that the decays of the
hybrid meson to a pair of ground state $1S$ conventional mesons
are suppressed~\cite{isgur,closeao}.

Unlike the above models, Qiao proposed that the $\y$ might be a
baryonium, containing charms, configured by the
$\Lambda_c$-$\overline{\Lambda}_c$~\cite{qiaocf}. This provides a
natural explanation for the absence of $\jpsi K \overline{K}$ and
$\ddb$ in its decays. Moreover, this scheme predicts
 \beq
 \frac{\Gamma [\y\to \piz \piz \jpsi]}
 {\Gamma [\y\to \pip \pim \jpsi]} \approx 1, \label{qiaoppjp}
 \eeq
which can be tested by the experiments.

Besides the fore-mentioned various interpretations, there is
another scheme which suggests that the $Y(4260)$ be a
$\chi_c$-$\rho$ molecule~\cite{liux}. The authors qualitatively
explained that the decay rate of $Y(4260)\to \pip \pim J/\psi$ is
greater than $\y \to \ddb$, and pointed out that $\y \to \piz\piz
J/\psi$ must be suppressed since $\rho^0$ only decays into $\pip
\pim$ but not $\piz\piz$.

In this Letter, we propose the $\y$ as a bound state composed of
the vector meson $\omega(783)$ and the $P$-wave charmonium state
$\chico(3510)$. In this scenario, we discuss its decay into
$\pi\pi\jpsi$, and expect that it decays into $\ppp\chico$ with
considerably large rate. The search for the latter is
experimentally reachable using the available data from the
$B$-factories. In addition, based on our scenario, we present some
predictions which are distinctive from those of other models.

\section{$\y$ as $\oc$ molecular state}

Since the $\y$ decays into $\psipp$, it is very natural to
consider that there is $c\bar{c}$ content in its wave function. We
try to find a narrow charmonium state and a narrow light meson to
form a $\jpc=1^{--}$ state, with the sum of their masses slightly
above the mass of the $\y$. There are not many such combinations,
and we find that the one consisting of a $1^{--}$ state $\omega$
(mass $782.59$~MeV/$c^2$~\cite{pdg}) and a $1^{++}$ state $\chico$
(mass $3510.59$~MeV/$c^2$~\cite{pdg}) satisfies the criteria. The
sum of the masses, $4293~\hbox{~MeV}/c^2$, is higher than the mass
of the $\y$ by 34~MeV/$c^2$, which is considered as the binding
energy between the two constituents to form the bound state. The
orbital angular momentum between $\omega$ and $\chico$ can be zero
to get the quantum number $\jpc=1^{--}$. In contrast to the
proposal of Ref.~\cite{liux}, here the $\y$ is an isoscalar
particle and has no isospin partner.

\section{Decays of the $\y$}

\begin{figure}
\begin{center}
\begin{minipage}{3.65cm}
\includegraphics[height=2.0cm]{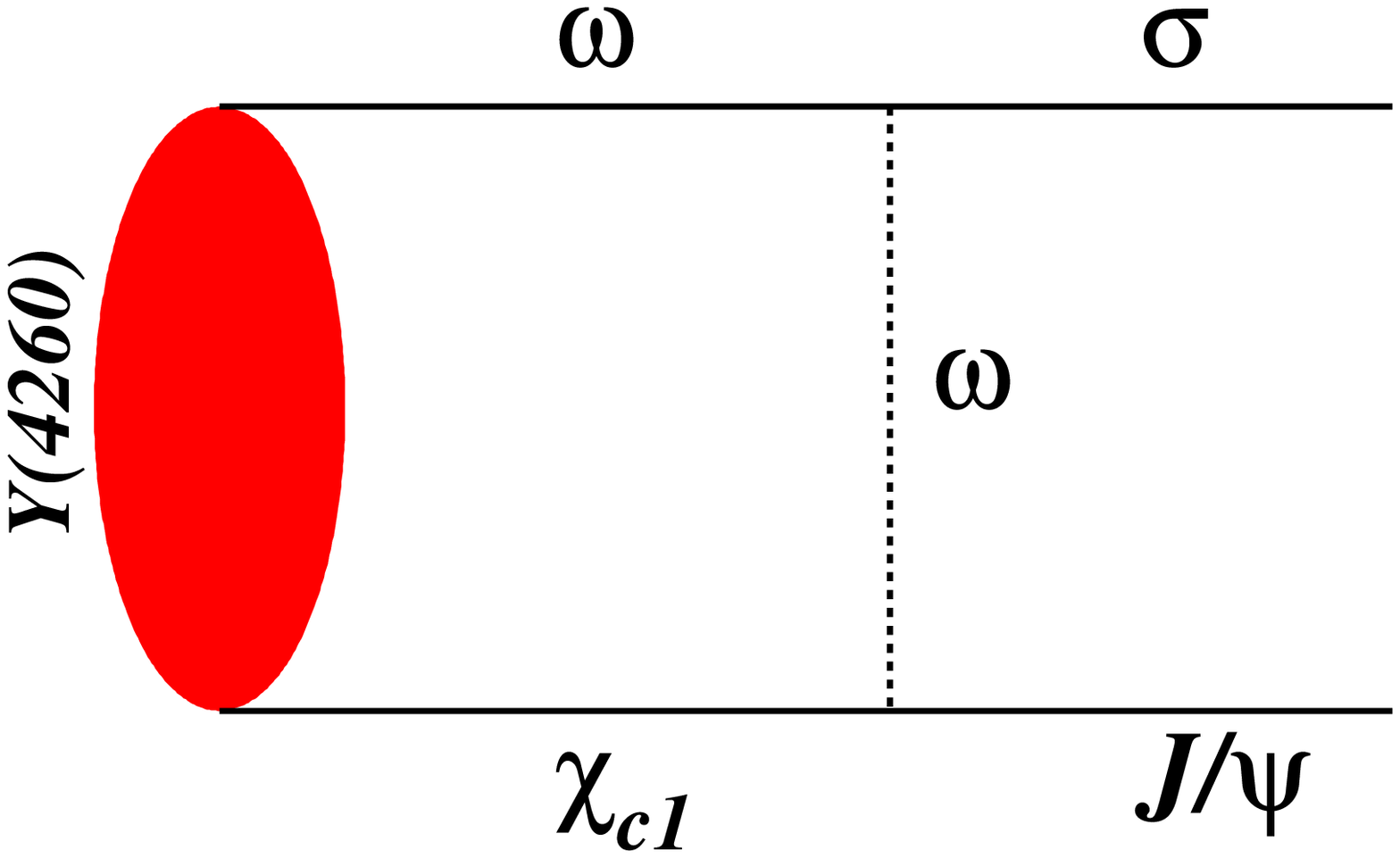}
\bigskip
\centerline{(a) $\omega$ exchange}
\end{minipage}
\begin{minipage}{3.65cm}
\includegraphics[height=2.0cm]{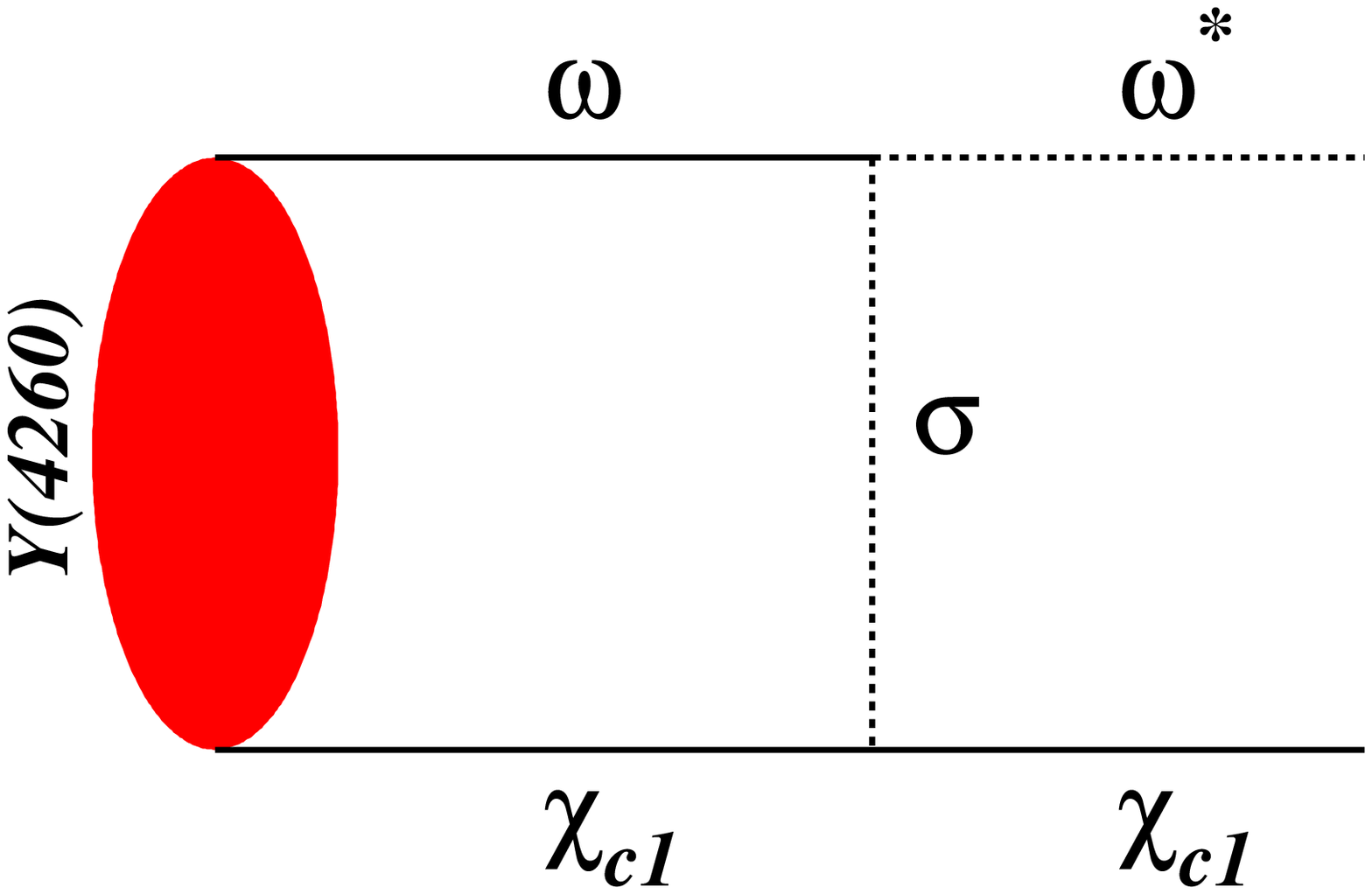}
\bigskip
\centerline{(b) $\sigma$ exchange}
\end{minipage}
\caption{\label{decay} Decay mechanism of the $\y$ into a
charmonium together with light hadrons.}
\end{center}
\end{figure}

The decay of the $\y$ to the observed $\ppjpsi$ mode is
illustrated in Fig.~\ref{decay}(a). In this picture, a virtual
$\omega$ is exchanged between the two bound constituents, and a
scalar particle like $\sigma$ or $f_0(980)$, and a vector
charmonium like $\jpsi$, $\psp$, or $\pspp$ are produced. This
decay mechanism can be verified by looking at the $\pp$ invariant
mass distribution in the BABAR data~\cite{babay4260}, which shows
signature of $\sigma$ at low mass side and of $f_0(980)$ at higher
mass side. In this scenario, according to the isospin symmetry, we
expect
 \beq
 \frac{\Gamma [\y\to \piz \piz \jpsi]}
 {\Gamma [\y\to \pip \pim \jpsi]}
 \approx 0.5, \label{ourppjp}
 \eeq
which is different from the predictions of being 1 in
Ref.~\cite{qiaocf} (Eq.~\eref{qiaoppjp}) and being 0 in
Ref.~\cite{liux}. This provides a proof to our scenario.

The $\y$ also decays via the exchange of a scalar particle as
illustrated in Fig.~\ref{decay}(b). In this case, the $\omega$
inside the $\y$ emits a scalar and turns into a photon or a
virtual $\omega$, which goes to $\ppp$ in final states; while the
$\chico$ inside the $\y$, after absorbing the virtual scalar,
becomes a real $\chico$ particle, which decays to $\gamma\jpsi$
with about 30\% branching fraction~\cite{pdg}. So a search for the
$\y$ decays into $\gamma \chico\to \gamma\gamma \jpsi$, or
$\omega^* \chico\to \ppp\gamma\jpsi$, provides another test of our
scenario.

\begin{figure}
\begin{center}
\includegraphics[height=3.0cm]{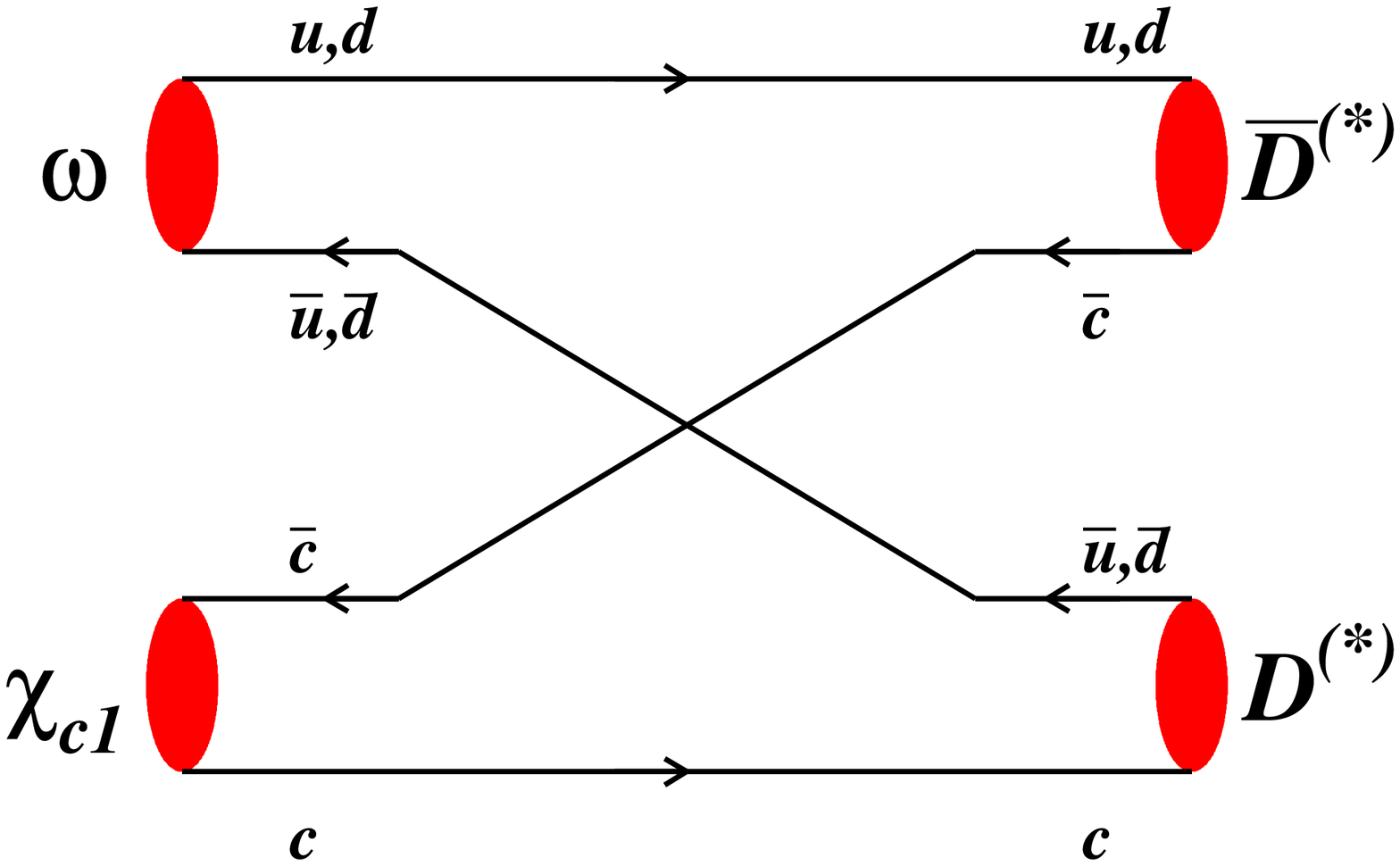}
\caption{\label{ddb} Decay mechanism for $\y\to
D^{(*)}\overline{D}^{(*)}$ final states.}
\end{center}
\end{figure}

The exchanges of light hadrons between the $\omega$ and $\chico$
inside the molecular state shown in Fig.~\ref{decay} require the 
exchange of at least two or three gluons in the quark level and
thus the processes are OZI suppressed~\cite{ozi_rule}, which in
principle, will be smaller than the final states produced by
changing quarks in the two constituents in the initial states
directly, as shown in Fig.~\ref{ddb}. However, in these decays of
the $\y$ into $D^{(*)}\overline{D}^{(*)}$ (which indicates
possible combinations such as $\ddb$, $D^* \overline{D}$,
$D\overline{D}^*$, and $D^*\overline{D}^*$), the rate is
suppressed due to color reconnection. The decays of the $\y$ into
charmed meson-pair with strange quarks have even lower rates since
a strange quark pair must be created.

From the above discussions, we see that the two decay mechanisms,
as shown in Figs.~\ref{decay} and \ref{ddb}, may have comparable
decay rates. The decays of the $\y$ into a charmonium state
together with a photon or some light mesons like pions will be
better tagging modes due to the clear signature and clean
environment; while the decays of the $\y$ to the charmed mesons,
have not been discovered experimentally by now because of the small
branching fractions of the D decay modes.

\section{Production of the $\y$}

\begin{figure}[htb]
\begin{center}
\includegraphics[height=3.0cm]{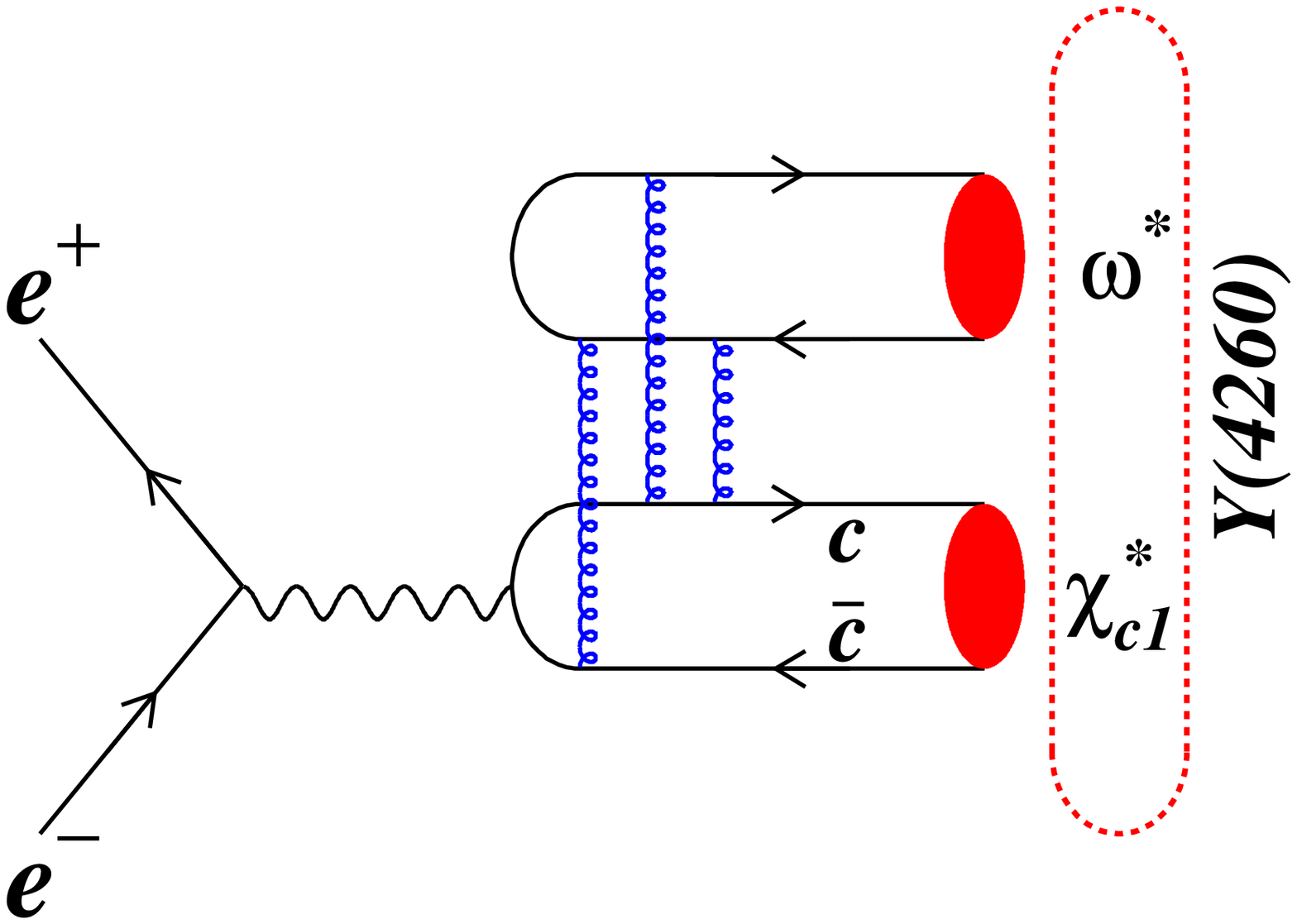}
\caption{\label{figprd} Production mechanism of the $\y$ in $\EE$
annihilation. The particle with ``$*$'' is virtual.}
\end{center}
\end{figure}

The production of the $\y$ in $\EE$ collision occurs via the
so-called hairpin mechanism~\cite{hairpin} which is shown in
Fig.~\ref{figprd}. Since the $\omega$ is produced from the gluons
emitted by the $c\overline{c}$ quarks, the production rate is
small relative to the $\psi$ states above 4~GeV/$c^2$. The
$\omega$ may be produced from a photon emitted from $\ccb$ quarks
too, but this amplitude is even smaller since it is proportional
to the QED fine structure constant $\alpha$.

\begin{figure*}[htb]
\begin{center}
\begin{minipage}{7.cm}
\includegraphics[height=4.5cm]{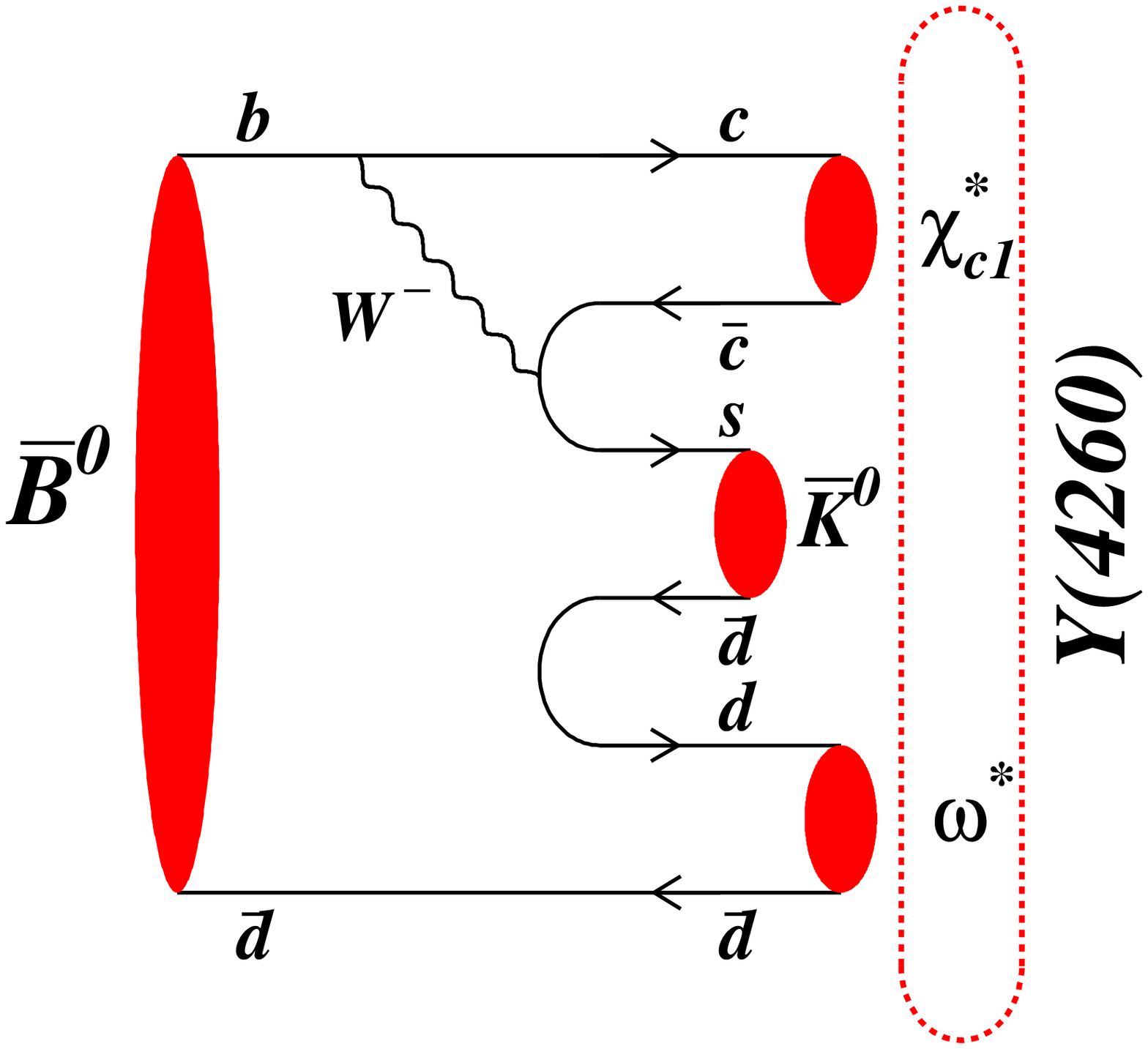}
\bigskip
\centerline{(a)}
\end{minipage}
\begin{minipage}{7.cm}
\includegraphics[height=4.5cm]{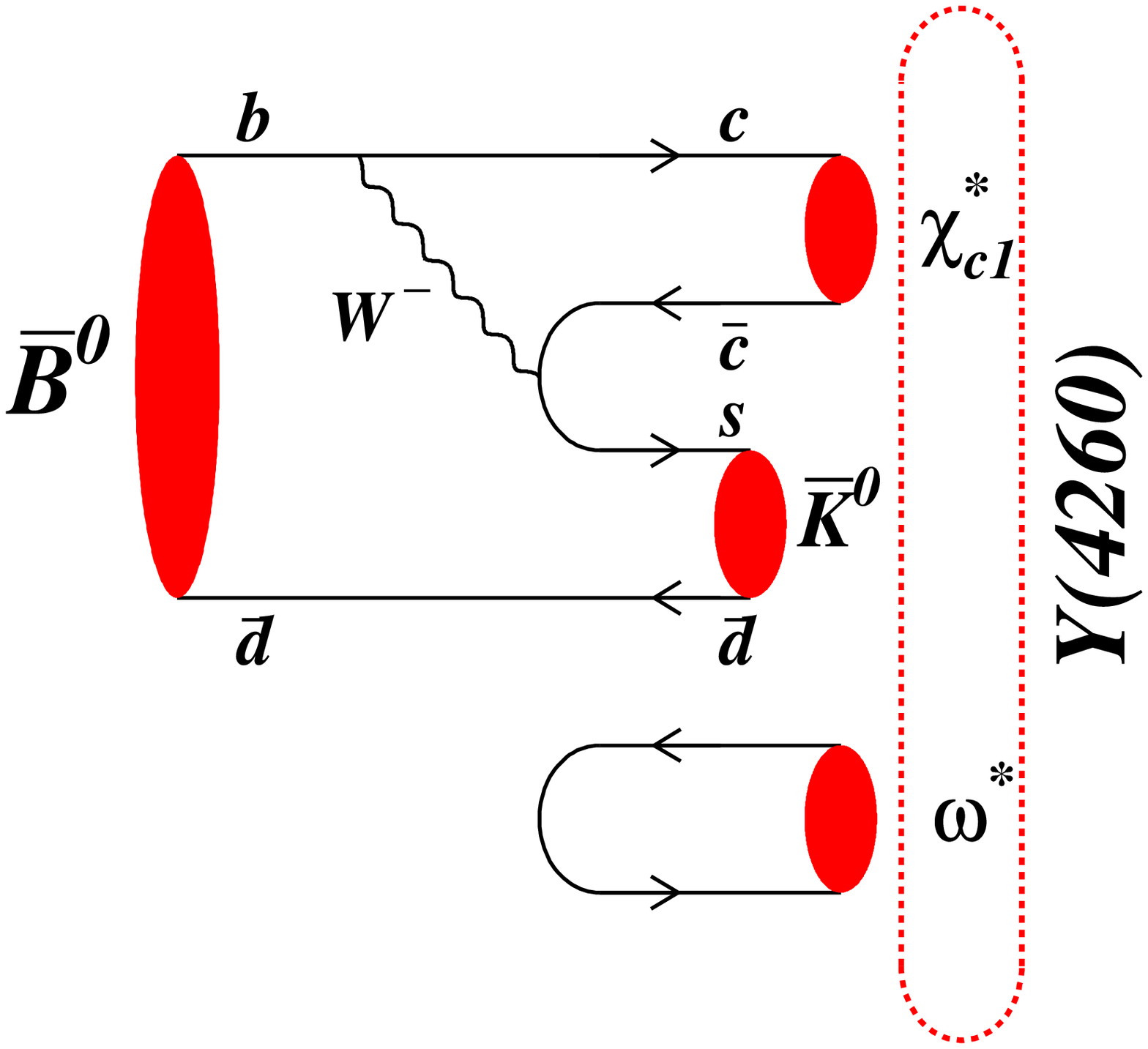}
\bigskip
\centerline{(b)}
\end{minipage}
\begin{minipage}{7.cm}
\includegraphics[height=4.5cm]{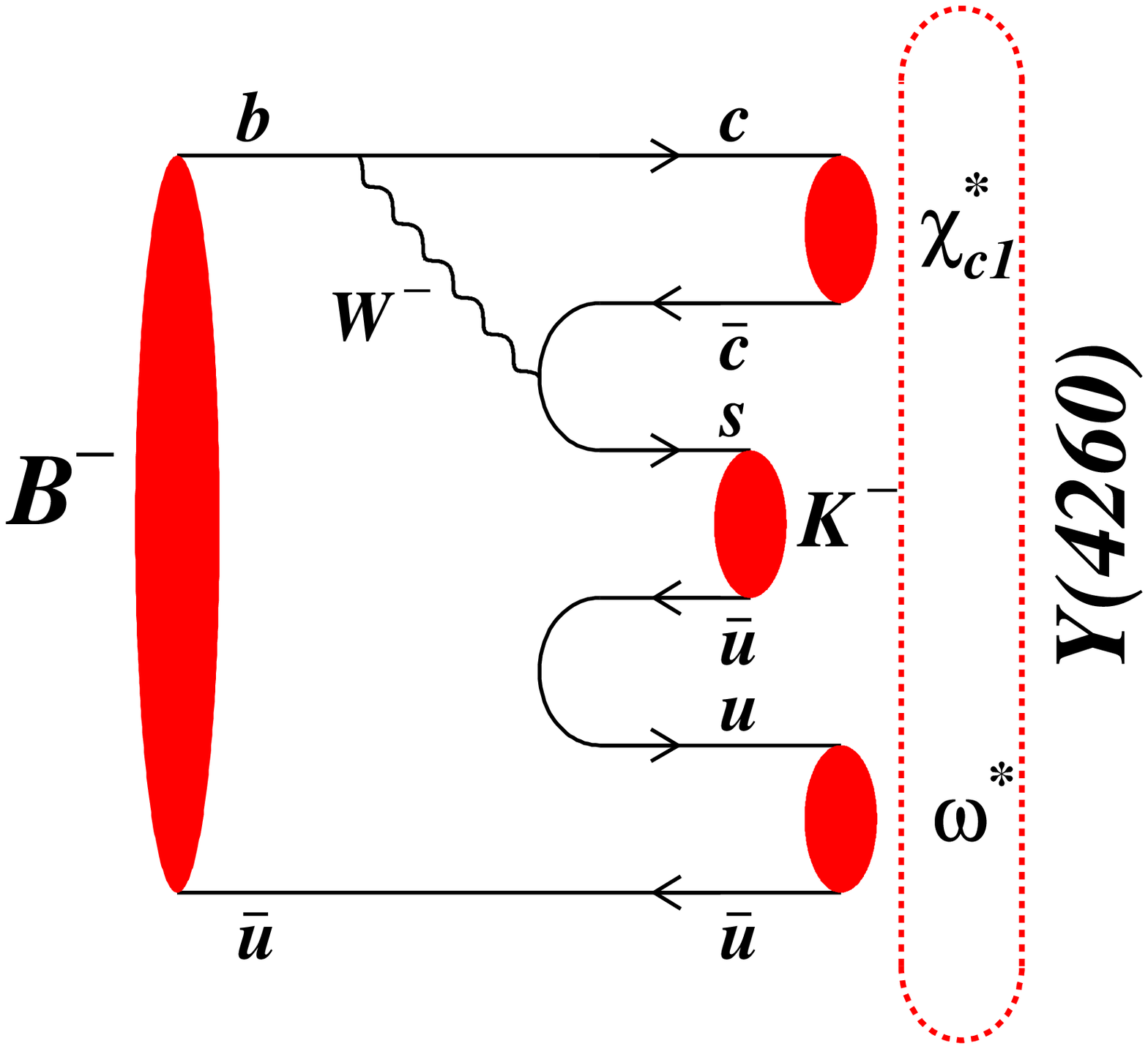}
\bigskip
\centerline{(c)}
\end{minipage}
\begin{minipage}{7.cm}
\includegraphics[height=4.5cm]{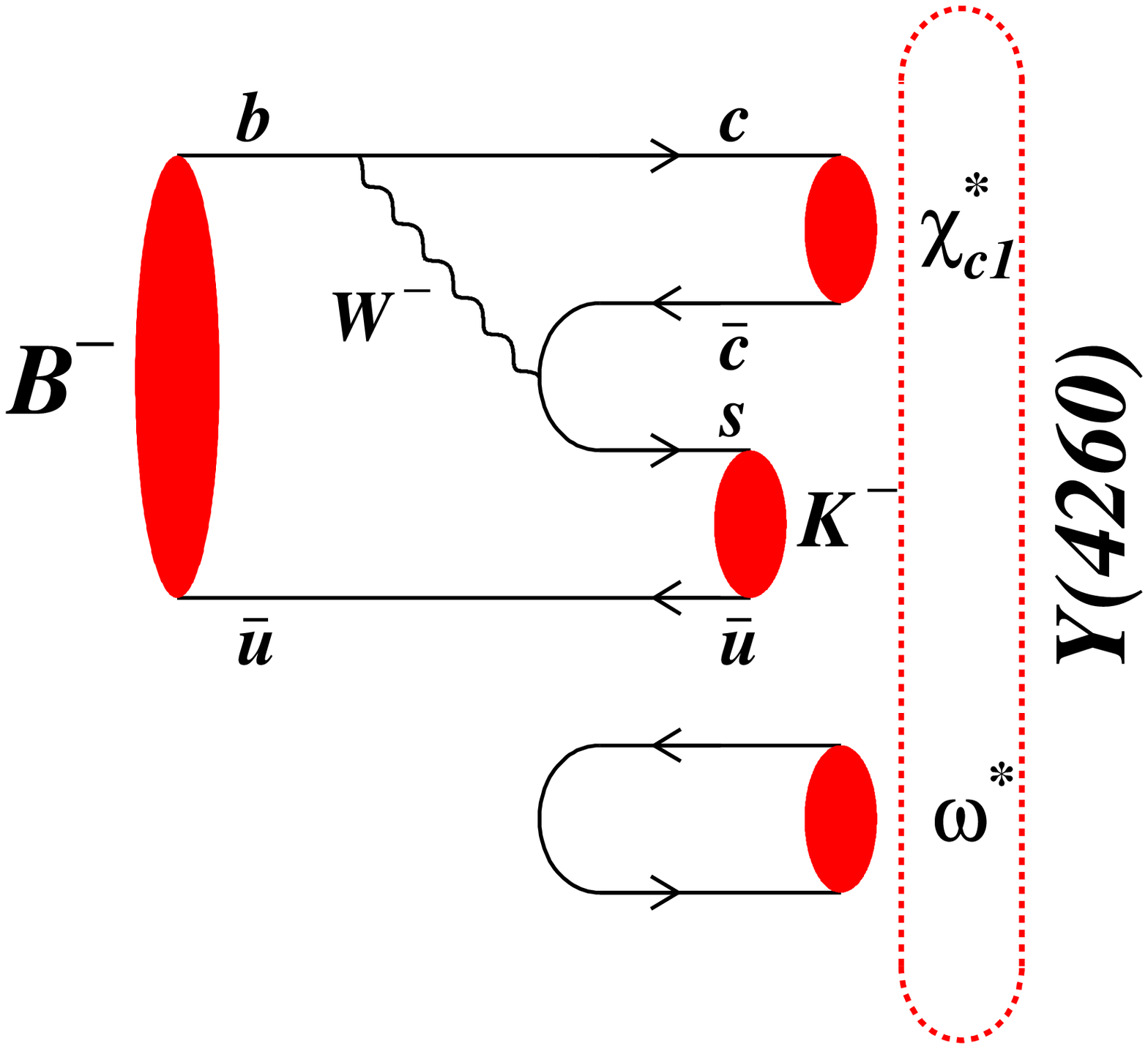}
\bigskip
\centerline{(d)}
\end{minipage}
\caption{\label{hairpinB} Production mechanism of the $\y$ in $B$
decays. (a) and (b) are for $\overline{B}^0$ decays, and (c) and (d) are 
for $B^-$ decays. The particle with ``$*$'' is virtual.}
\end{center}
\end{figure*}

\begin{figure*}
\begin{center}
\begin{minipage}{7.cm}
\includegraphics[height=3.0cm]{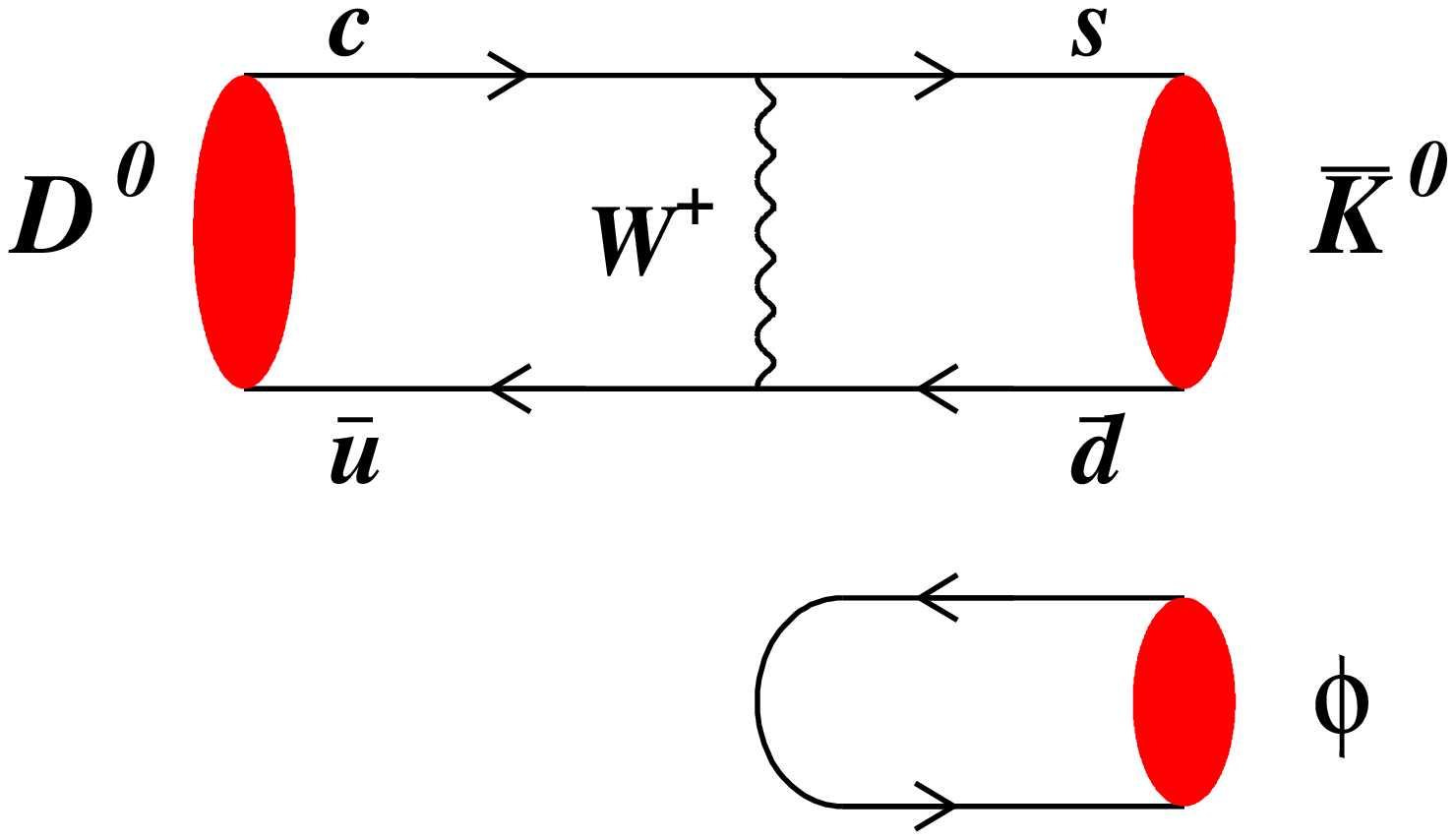}
\bigskip
\centerline{(a)}
\end{minipage}
\begin{minipage}{7.cm}
\includegraphics[height=3.0cm]{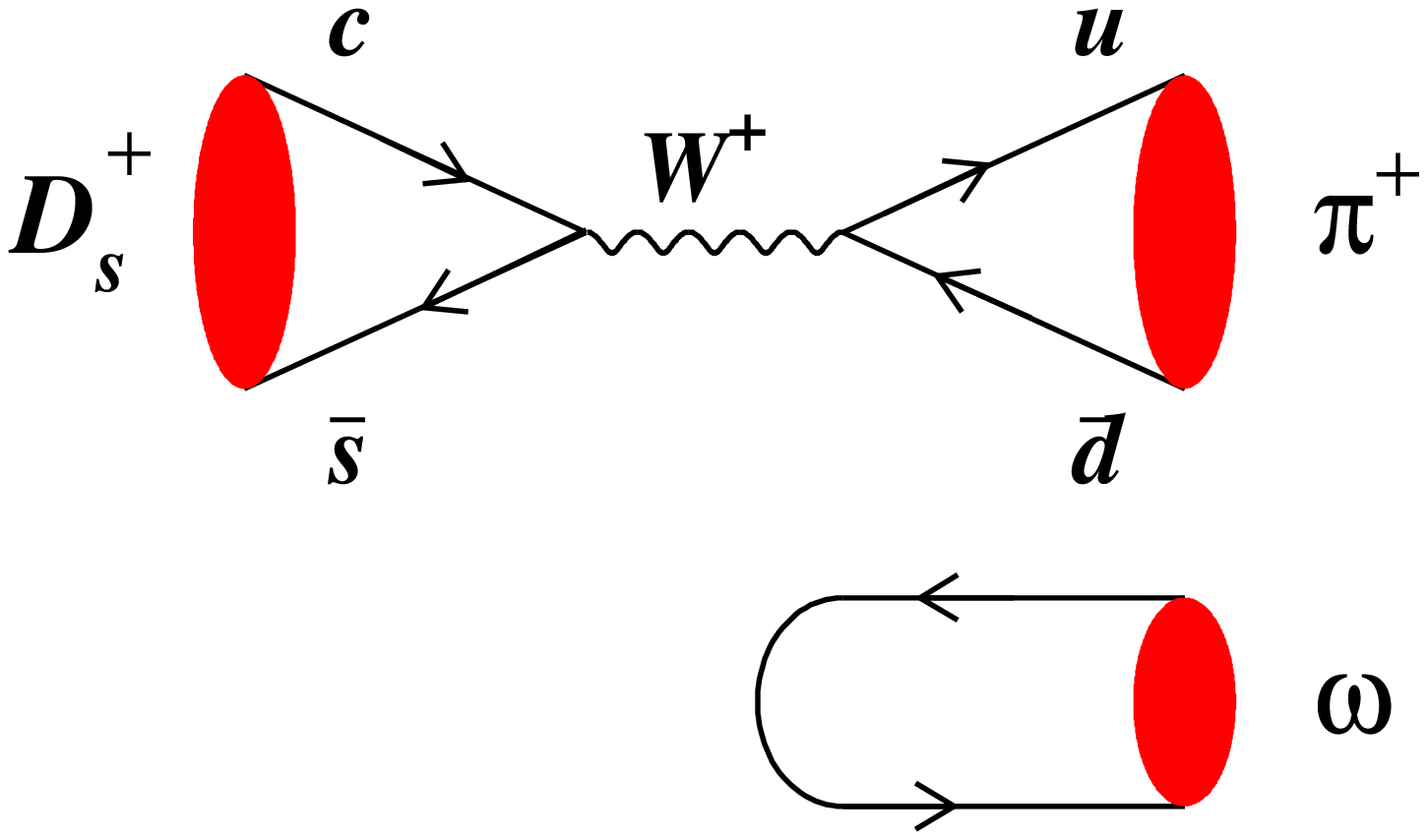}
\bigskip
\centerline{(b)}
\end{minipage}
\caption{\label{hairpinD} Hairpin diagrams for
$D^0 \to \overline{K}^0 \phi$ (a) and
$D_s^+ \to \pi^+ \omega$ (b) decays.}
\end{center}
\end{figure*}

The $\y$ was also found in $B$ decays in association with a $K$
meson~\cite{babajp}. It may be produced via a spectator diagram,
as shown in Fig.~\ref{hairpinB}(a) and (c), and via a hairpin
diagram, as shown in Fig.~\ref{hairpinB}(b) and (d). Although in
the hairpin mechanism, the $\omega$ is produced from the gluons
emitted by the quarks, so its amplitude is thought to be
suppressed, but there is indication from the weak decays of charmed
mesons that such suppression is not severe: the experiments
measured~\cite{pdg} \( \BR(D^0\ra \phi \overline{K^0})=(9.4\pm
1.1) \times 10^{-3} \) and \(\BR(D^+_s \ra \omega \pi^+)=(2.8\pm
1.1) \times 10^{-3} \) can be explained by the hairpin
diagrams~\cite{wangli} shown in Fig.~\ref{hairpinD}. They are
smaller by only about an order of magnitude compared to the
signature modes which go through spectator diagrams without color
reconnection: $ \BR(D^0 \ra \rho^+ K^-)=(10.1\pm 0.8) \% $ and $
\BR(D^+_s \ra \phi \pi^+)=(3.6\pm 0.9) \%$~\cite{pdg}. If this is
extended to the weak decays of $B$ mesons, one may expect that the
diagrams of Fig.~\ref{hairpinB}(b) and \ref{hairpinB}(d) be
important in the production of the $\y$ in $B$ decays.

\section{Discussions}

Although the above discussions are limited to the $\oc$ bound
state, the scenario can be naturally extended to other bound
states consisting of a light meson and a charmonium state. If the
bound state of $\chico$ and $\omega$ exists, by the same
mechanism, a $\chi_{c0}$ or a $\chi_{c2}$ can also form a bound
state with an $\omega$. At the same time, besides $\omega$, other
$SU(3)$ singlet light hadrons, like $\phi$, $\eta$, and
$\eta^\prime$, can also form bound states with charmonia, such as
$\jpsi$, $\psp$, $\chicz$, $\chico$, $\chict$ and $h_c$.
Table~\ref{more_states} gives the sum of the masses of a light
hadron ($\eta$, $\eta^\prime$, $\omega$ and $\phi$) and a
charmonium state ($\jpsi$, $\chicz$, $\chico$, $\chict$ and
$h_c$). Considering a binding energy of a few to a few ten MeV in
forming the bound state, we can see the newly observed states
$X(3872)$~\cite{bellex3872} could be interpreted as an
$\omega\jpsi$ bound state, the $Y(3940)$~\cite{belley3940} could
be an $\eta\chicz$ bound state. Furthermore, there are many other
possible combinations which have no experimental evidence yet.
Their decay properties can be analyzed in the same way as in this
Letter, and the production of these bound states in $\EE$
collision (if it is a $\jpc=1^{--}$ state) and in $B$ decays
follows the similar mechanisms described in previous sections.
These bound states should be searched for by the $B$-factories
both in ISR data and in $B$ decays.

\btbl[hbtp] \caption{\label{more_states} The sum of the masses (in
MeV/$c^2$) of a light meson and a charmonium state. A bound state
of each possible combination could be produced by emitting a few
or a few ten MeV binding energy. The numbers underlined may
correspond to the states which have been observed experimentally.}
\begin{center}
\btbu{c| c c  c c c} \hline\hline
          & $\jpsi$ & $\chicz$ & $\chico$ & $\chict$ & $h_c$ \\\hline
 $\eta$         &  3645 & \underline{3963}
                                     & 4058   & 4104 & 4073 \\
 $\eta^\prime$  &  4055 & 4373   & 4468   & 4514 & 4483 \\
 $\omega$       &  \underline{3880} & 4198
                           & \underline{4293}   & 4339 & 4308 \\
 $\phi$         &  4116 & 4435   & 4530   & 4576 & 4544 \\
\hline\hline \etbu
\end{center}
\end{table}

\section{Summary}

The $\y$ observed by the \bbcol is proposed as an $\oc$ molecular
state, which may decay into a charmonium state together with some
light particles, with comparable rates as for the decays to open
charm final states. A few predictions are made and could be tested
in the $B$-factories. The decays of $\y\to\ppp\chico$ or
$\gamma\chico$ should be searched for in high priority.

\section*{Acknowledgments}

The paper was motivated in the informal workshop on ``New Hadron
States" in Peking University in October 2005. We acknowledge the
helpful discussions with many participants. We also express thanks to
Dr. H.B. Li for useful discussion.

\end{document}